\documentclass[12pt]{article}
\usepackage{amsthm,amsmath,amsfonts,amssymb,cases}
\usepackage{color}

\newtheorem{lemma}{Lemma}
\newtheorem{theorem}[lemma]{Theorem}

\begin{document}
\title{Ground state properties in non-relativistic  QED}
\author{\vspace{5pt} D. Hasler$^1$\footnote{
E-mail: david.hasler@math.lmu.de \quad
On leave from College of William \& Mary} and I.
Herbst$^2$\footnote{E-mail: iwh@virginia.edu.} \\
\vspace{-4pt} \small{$1.$ Department of Mathematics,
Ludwig Maximilians University} \\ \small{Munich, Germany }\\
\vspace{-4pt}
\small{$2.$ Department of Mathematics, University of Virginia,} \\
\small{Charlottesville, VA, USA}\\}
\date{}
\maketitle

\begin{abstract}
We discuss recent  results concerning the ground state of   non-relativistic
quantum electrodynamics  as a function of a magnetic coupling constant
or the fine structure constant, obtained by the authors in   \cite{HH10-1,HH10-2,HH10-3}. 
\end{abstract}


\section{Introduction}

We consider a system of finitely many non-relativistic quantum mechanical 
electrons bound to a static nucleus. The electrons are minimally coupled to the
quantized electromagnetic field, and we denote the   coupling constant  by $g$.
We impose an ultraviolet cutoff on the electromagnetic vector potential
appearing in the covariant  derivatives.

Models of this type are known  as non-relativistic quantum electrodynamics (qed).
They provide a  reasonable description of 
microscopic low energy phenomena   involving electrons, nuclei, and
photons.
A systematic mathematical investigation of
these models started in the mid 90s with the work of V.
Bach, J. Fr\"ohlich, and I.M. Sigal  \cite{BFS98,BFS98-2,BFS99}.
They showed   existence of  ground states. 
Furthermore, they showed that
excited bound states of the unperturbed system  become unstable 
and turn into resonances when the electrons are coupled to the radiation field.
To prove  this result they introduced an operator theoretic 
renormalization analysis. Later,  the existence of ground states was shown in more 
generality 
by M. Griesemer, E.H. Lieb, and M. Loss, see \cite{GLL01,LL03}. 

In \cite{HH10-2} we   showed   that the ground state of an atom with spinless electrons is an analytic 
function of the  coupling constant $g$.  That result is explained in 
Section \ref{sec:2}, and it provides an algorithm to determine 
the ground state to arbitrary precision. 
To obtain the result we used the  operator theoretic renormalization analysis of \cite{BFS98}
and that renormalization preserves anlyticity \cite{GH09}.

In Section \ref{sec:3}, we  consider expansions in the fine structure constant $\alpha$.  
We consider a scaling where the ultraviolet cutoff is   of the order of the binding energy 
of the unperturbed atom.  In this scaling  lifetimes of excited states 
of atoms were calculated which agree with experiment \cite{HHH08}. 
V. Bach, J. Fr\"ohlich, and 
A. Pizzo \cite{BFP06,BFP09}  showed that there exists an asymptotic expansion of the 
ground state and the ground state energy  with $\alpha$ dependent 
coefficients. 
 In \cite{HH10-2}  this result was extended and  it was shown  that these expansions
are  convergent. 
Furthermore, it was shown  in \cite{HH10-3} that the ground state energy as
well as the ground state  are $k$-times continuously differentiable  functions of $\alpha$
 respectively $\alpha^{1/2}$ on some nonempty $k$-dependent interval $[0,c_k)$.
This result implies that there 
are no logarithmic terms in this  scaling limit. This resolves an open issue raised in \cite{BFP09}, since 
for other  scalings of the ultraviolet cutoff  logarithmic terms do occur \cite{bet47,HS02,BCVV09}.

\section{Model and analyticity of the ground state}

\label{sec:2}

 We introduce the bosonic Fock space over the one photon Hilbert space 
$\mathfrak{h} := L^2( \mathbb{R}^3 \times \mathbb{Z}_2 )$ and set
$$
\mathcal{F} := \mathbb{C} \oplus \bigoplus_{n=1}^\infty  S_n ( \mathfrak{h}^{\otimes^n}  ) ,
$$
where   $S_n$  denotes the orthogonal projection onto the subspace of totally symmetric
tensors in  $\mathfrak{h}^{\otimes^n}$.
By  $a^*(k,\lambda)$ and $a(k,\lambda)$, with $(k,\lambda) \in \mathbb{R}^3 \times \mathbb{Z}_2$,  we denote the  so called creation and annihilation operator. They 
 satisfy the
following commutation relations, which are to be understood in the sense of distributions,
$$
[a(k,\lambda), a^*(k',\lambda') ] = \delta_{\lambda \lambda'} \delta(k - k')   , \qquad [a^\#(k,\lambda), a^\#(k',{\lambda'}) ]  = 0 \; ,
$$
where $a^{\#}$ stands for  $a$ or $a^*$. The operator $a(k,\lambda)$ annihilates the vacuum
$(1,0,...) \in \mathcal{F}$.
We define the operator of the free field energy
by
$$
H_f := \sum_{\lambda=1,2} \int  a^*(k,\lambda)   |k|  a(k,\lambda) d^3k .
$$
For $\lambda=1,2$ we introduce the so called polarization vectors
$
\varepsilon(\cdot , \lambda) : S^2 := \{ k \in \mathbb{R}^3 | |k| = 1 \} \to \mathbb{R}^3
$
to be  maps such that for each $k \in S^2$ the vectors $\varepsilon(k,1), \varepsilon(k,2),k$ form an orthonormal basis of $\mathbb{R}^3$. 
For $x \in \mathbb{R}^3$ we define the field operator
\begin{equation} \label{eq:afield}
A_{\Lambda}(x) :=  \sum_{\lambda=1,2} \int_{|k| < \Lambda} \frac{d^3k }{\sqrt{2 |k|}} \left[ e^{-ik \cdot x}
\varepsilon(\widehat{k},\lambda) a^*(k,\lambda)  + e^{ik \cdot x} \varepsilon( \widehat{k}, \lambda) a(k, \lambda) \right] \ ,
\end{equation}
where 
$0 < \Lambda $ is a finite  ultraviolet cutoff and $\widehat{k} := k /|k|$.
The Hilbert space is $\mathcal{H} := \mathcal{H}_{\rm at} \otimes \mathcal{F}$, where 
$$
\mathcal{H}_{\rm at}  := \bigwedge^N  L^2(\mathbb{R}^{3}) 
$$
is the  Hilbert space  describing  $N$ spin-less electrons.
We study  the following operator in $\mathcal{H}$
\begin{equation} \label{eq:hamiltoniandefinition}
H_{g}  :=  \ \  \sum_{j=1}^N  ( p_j + g A_\Lambda( x_j) )^2   +  V   + H_f ,
\end{equation}
where $x_j \in \mathbb{R}^3$ denotes the coordinate of the $j$-th electron,
$p_j = - i \partial_{x_j}$, and 
$V$ denotes the potential. 
For  the result concerning analyticity in the coupling constant $g$ on a disk 
$D_{r} := \{ z \in \mathbb{C} | |z | < r \}$, we need the following hypothesis.
It  contains assumptions about  the atomic Hamiltonian 
$H_{\rm at} :=   \sum_{j=1}^N p_j^2   + V $ acting in  $\mathcal{H}_{\rm at}$.

\vspace{0.3cm}
\noindent
{\bf Hypothesis (H)} The potential $V$ satisfies the following properties:
\begin{itemize}
\item[(i)] $V$ is invariant  under permutations and  rotations.
\item[(ii)] $V$ is infinitesimally operator bounded with respect to $\sum_{j=1}^N p_j^2$.
\item[(iii)] $E_{\rm at} := {\rm inf}\, \sigma(H_{\rm at} )$ is a non-degenerate isolated eigenvalue of $H_{\rm at}$.
\end{itemize}
All assumptions of 
Hypothesis  (H) are satisfied for the hydrogen atom. Part (i) is satisfied for 
atoms, but not for molecules with static nuclei. We note that  (iii)  is a restrictive assumption.  

\begin{theorem} \label{thm:main0} Suppose  (H). Then there exists
a positive constant $g_0$ such that for all $g \in D_{g_0}$
the operator $H_{g}$ has a non-degenerate eigenvalue $E(g)$ with eigenvector $\psi(g)$ and eigen-projection
$P(g)$ satisfying the following properties.
\begin{itemize}
\item[(i)] For $g \in \mathbb{R} \cap D_{g_0}$, $E_{}(g) = {\rm inf} \sigma ( H_{g}) $. 
\item[(ii)] $g \mapsto E_{}(g)$ and $g \mapsto \psi_{}(g)$
are analytic on $D_{g_0}$.
\item[(iii)] $g \mapsto P_{}(g)$ is analytic on $D_{g_0}$ and
$P_{}(g)^* = P_{}(\overline{g})$.
\end{itemize}
\end{theorem}

Concerning the proof of the   theorem, we note that 
 the  ground state energy is embedded in continuous
spectrum.  In such a situation  analytic perturbation theory is
typically not applicable and other methods have to be employed.
In \cite{HH10-2}   Theorem \ref{thm:main0} is proven using 
   a variant of the operator theoretic renormalization analysis.
Using the rotation invariance assumption 
of  Hypothesis (H) one can prove 
that marginal terms in the renormalization analysis are absent.
This implies that the  renormalization analysis converges.
Theorem \ref{thm:main0} can then be shown  using that  renormalization  preserves analyticity \cite{HH10-1,GH09}.

Theorem \ref{thm:main0}  implies that the ground state and the ground state energy 
admit  convergent power series expansions in $g$. The coefficients of these 
expansions can be calculated by means of analytic perturbation theory.
To this end, one introduces an infrared  cutoff which renders 
all expansion coefficients finite. In \cite{HH10-2} it was shown 
using  a continuity argument, that  the individual expansion coefficients converge as the infrared 
cutoff is removed. 
This is not obvious;   the expansion coefficients  obtained by regular perturbation theory, \cite{K},
can involve cancellations of  infrared divergent terms  \cite{HH10-1}.

\section{Expansions in  the fine structure constant}

\label{sec:3}
In this section, we consider the ground state and the ground state energy
of a hydrogen atom as a function of the fine structure constant $\alpha$. We assume that the ultraviolet cutoff
is of the order of the binding energy of the unperturbed atom. In suitable units
the corresponding Hamiltonian is 
$$
H_{\alpha, \Lambda} := ( p +    \alpha^{3/2} A_\Lambda( \alpha x) )^2   
- \frac{1}{|x|}    + H_f .
$$
By a scaling transformation we can relate this  operator to the 
operator 
$$
\widetilde{H}_{\alpha,\Lambda}  :=  
 ( p+ \sqrt{\alpha} A_\Lambda( x) )^2    - \frac{\alpha}{|x|}       + H_f 
$$
using the following unitary equivalence
$\widetilde{H}_{\alpha, \alpha^2 \Lambda}
 \cong \alpha^2 H_{ \alpha, \Lambda}$.
We are interested in the behavior of the ground state and the ground state energy
of  $H_{\alpha,\Lambda}$ as $\alpha \downarrow 0$ while $\Lambda$ remains constant.
The  ground state and the 
ground state energy are smooth in the sense of the following theorem \cite{HH10-3}.

\begin{theorem} \label{cor:main1}  Suppose (H) and let $\Lambda > 0$. 
There exists a positive  $\alpha_0$ such that for $\alpha \in [ 0 , \alpha_0)$ the operator
$H_{\alpha,\Lambda}$ has a ground state $\psi(\alpha^{1/2})$ with ground
state energy $E(\alpha)$ such that we have the convergent expansions on $[0,\alpha_0)$
\begin{equation} \label{eq:expansion}
E(\alpha) = \sum_{n=0}^\infty E^{(2n)}_\alpha \alpha^{3 n }  ,\qquad
\psi(\alpha^{1/2}) = \sum_{n=0}^\infty \psi_\alpha^{(n)} \alpha^{3n/2} .
\end{equation}
The coefficients   $E_\alpha^{(n)}$ and $\psi_\alpha^{(n)}$ are as functions of $\alpha$
in $C^\infty([0,\infty))$ and $C^\infty([0,\infty);\mathcal{H})$, respectively.
For every $k \in \mathbb{N}_0$ there exists a positive  $\alpha_0^{(k)}$ such that
$\psi(\cdot)$ and $E(\cdot)$ are $k$-times continuously differentiable on $[0,\alpha_0^{(k)})$.
\end{theorem}
By the differentiability property of
Theorem  \ref{cor:main1} and Taylor's theorem one can write 
the ground state and the ground state energy  in terms of an asymptotic
series with constant coefficients in the sense of \cite{reesim4}.
To  prove  Theorem  \ref{cor:main1}, we consider the Hamiltonian
$$
H(g, \beta,\Lambda) := ( p +  g A_\Lambda( \beta x) )^2   -  \frac{1}{|x|}  + H_f .
$$
Using the identity  $H(\alpha^{3/2},\alpha,\Lambda) = H_{\alpha,\Lambda}$,
 Theorem  \ref{cor:main1}  will follow as an application of Theorem \ref{thm:main1}, below.
A corollary of that theorem is that the ground state of $H(g,\beta,\Lambda)$ is analytic in $g$ with
coefficients which are $C^\infty$ functions of $\beta$. To state the theorem precisely,  let  $X$ be a Banach space and let  $C^k_B(\mathbb{R};X)$ denote 
the  space  of $X$-valued 
functions having bounded, continuous derivatives up to order $k$  normed by 
$
\| f \|_{C^k_B(\mathbb{R};X)} := \max_{0 \leq s \leq k} \sup_{x \in \mathbb{R}} \| D_x^s f(x) \|_X .
$
\begin{theorem} \label{thm:main1}  Suppose (H), let $k \in \mathbb{N}_0$, and $\Lambda > 0$. Then there exists
a positive $g_0$ such that for all  $(g,\beta)  \in D_{g_0} \times \mathbb{R}$
the operator $H(g,\beta,\Lambda)$ has an eigenvalue $E_{\beta}(g)$ with eigenvector $\psi_{\beta}(g)$ and eigen-projection
$P_{\beta}(g)$ satisfying the following properties.
\begin{itemize}
\item[(i)] For $g \in \mathbb{R} \cap D_{g_0}$ we have  $E_{\beta}(g) = {\rm inf}  \sigma ( H_{g,\beta}) $, and
 for all $g \in D_{g_0}$ we have $P_{\beta}(g)^* = P_{\beta}(\overline{g})$.
\item[(ii)] $g \mapsto  E_{(\cdot)}(g)$,  $g \mapsto  \psi_{(\cdot)}(g)$, and
 $g \mapsto  P_{(\cdot)}(g)$ are analytic  functions on $D_{g_0}$ with values in $C_B^k(\mathbb{R})$,
 $C^k_B(\mathbb{R};\mathcal{H})$, and $C^k_B(\mathbb{R};\mathcal{B}(\mathcal{H}))$, respectively.
\end{itemize}
\end{theorem}
In \cite{HH10-3} Theorem \ref{thm:main1} is  shown using  an operator theoretic renormalization
analysis, which involves     controlling   arbitrary high derivatives with respect to $\beta$.


\begin{thebibliography}{30}


\bibitem{BFP06} V. Bach, J. Fr\"ohlich, A. Pizzo, {\it  Infrared-finite Algorithms in QED: the groundstate of an Atom interacting with the quantized radiation field},
Comm. Math. Phys. 264 (2006), no. 1, 145--165.

\bibitem{BFP09} V. Bach, J. Fr\"ohlich, A. Pizzo, {\it  Infrared-finite Algorithms in QED. II. The expansion of the groundstate of an Atom interacting with the quantized radiation field},
Adv. Math. 220 (2009), no. 4, 1023--1074.



\bibitem{BFS98}  V. Bach,  J. Fr\"ohlich, I.M. Sigal,
{\em Renormalization group analysis of spectral problems in quantum field theory},
 Adv. Math.  137 (1998),  205--298.




\bibitem{BFS98-2} V. Bach, J. Fr\"{o}hlich, I.M. Sigal,
{\em Quantum electrodynamics of confined nonrelativistic particles}.
Adv. Math. {  137} (1998), 299--395.

\bibitem{BFS99} V. Bach, J. Fr\"ohlich, I.M.  Sigal, {\it  Spectral analysis for systems of atoms and molecules coupled to the quantized radiation field},
 Comm. Math. Phys. 207 (1999), no. 2, 249--290.




\bibitem{BCVV09} J-M. Barbaroux, T. Chen, S. Vugalter, V. Vougalter
{\it  Quantitative estimates on the Hydrogen ground state energy in non-relativistic QED }. {\tt mp\_arc 09-48 }






\bibitem{bet47} H.A.~Bethe, {\em The Electromagnetic Shift of Eneregy Levels}, Phys. Rev.  72
(1947), 339--341.



\bibitem{GLL01}
M. Griesemer, E. Lieb, M. Loss, {\it
Ground states in non-relativistic quantum electrodynamics}.
Invent. Math. 145 (2001), no. 3, 557--595.



\bibitem{GH09} M. Griesemer, D. Hasler, {\em Analytic Perturbation Theory and Renormalization Analysis of Matter Coupled to Quantized Radiation},
Ann. Henri Poincar\'e.






\bibitem{HS02} C. Hainzl, R. Seiringer, {\it
Mass renormalization and energy level shift in non-relativistic QED},  Adv. Theor. Math. Phys.   6 (2002),   847--871.







\bibitem{HHH08} D. Hasler, I. Herbst, M. Huber, {\it
On the lifetime of quasi-stationary states in non-relativistic QED},
Ann. Henri Poincar\'e 9 (2008),  1005--1028.


\bibitem{HH10-1} D. Hasler, I. Herbst, {\it Ground state properties of the spin boson model},   submitted. {\tt  arXiv:1003.5923}

\bibitem{HH10-2} D. Hasler, I. Herbst, {\it Convergent expansions in non-relativistic QED: Analyticity of the ground state}  {\tt arXiv:1005.3522}

\bibitem{HH10-3} D. Hasler, I. Herbst, {\it Smoothness and analyticity of perturbation expansions in QED} {\tt arXiv:1005.3522}



\bibitem{LL03} E.H. Lieb, M. Loss, {\em Existence of atoms and molecules in non-relativistic quantum electrodynamics},  Adv. Theor. Math. Phys.  7  (2003),  no. 4, 667?710. 

\bibitem{K} T. Kato, \emph{Perturbation theory for linear operators}, Springer Verlag, New York, 1966, pp.75-80.

\bibitem{reesim4} M. Reed and B. Simon, {\it Methods of modern mathematical physics. IV. Analysis of operators},
Academic Press, New York-London, 1978.









\end{thebibliography}

\end{document}